\documentclass[aps,prd,preprintnumbers,showpacs,superscriptaddress,tighten,nofootinbib,twocolumn]{revtex4}
\usepackage{amssymb}
\usepackage{mathrsfs}
\usepackage{mathbbold}
\usepackage{graphicx}
\usepackage{amsmath}
\usepackage{epsfig}
\usepackage{mciteplus}
\usepackage{amsmath,color}
\usepackage{multirow}
\usepackage{rotating,tabularx}
\usepackage{lscape}
\usepackage{ulem}

\begin{document}

\title{Systematic search for successful lepton mixing
patterns with nonzero $\theta_{13}$}

\author{Werner Rodejohann}
\email{werner.rodejohann@mpi-hd.mpg.de}

\affiliation{Max-Planck-Institut f{\"u}r Kernphysik, Postfach
103980, 69029 Heidelberg, Germany}

\author{He Zhang}
\email{he.zhang@mpi-hd.mpg.de}

\affiliation{Max-Planck-Institut f{\"u}r Kernphysik, Postfach
103980, 69029 Heidelberg, Germany}

\author{Shun Zhou}
\email{zhoush@mppmu.mpg.de}

\affiliation{Max-Planck-Institut f{\"u}r Physik
(Werner-Heisenberg-Institut), F{\"o}hringer Ring 6, D-80805
M{\"u}nchen, Germany}


\preprint{MPP-2011-84}

\begin{abstract}
We perform a systematic search for simple but viable lepton mixing
patterns. Our main criterion is that the mixing matrix can be
parameterized by three rotation angles, which are simple fractions
of $\pi$. These simple rotation angles possess exact expressions for
their sines and cosines, and often arise in the flavor symmetry
models. All possible parameterizations of the mixing matrix are
taken into account. In total, twenty successful mixing patterns are
found to be consistent with the latest neutrino oscillation data
(including the recent T2K results) in the CP conserving case,
whereas fifteen mixing patterns are allowed in the maximal CP
violating case. Potential radiative corrections to the constant
mixing patterns are also calculated by solving the renormalization
group equations.
\end{abstract}

\begin{center}
\pacs{14.60.Pq, 14.60.Lm}
\end{center}

\maketitle

\section{Introduction}
\label{sec:intro}

Recent solar, atmospheric, reactor and accelerator neutrino
experiments have provided us with compelling evidence that neutrinos
are massive and lepton flavors are mixed. In the framework of
three-flavor neutrino oscillations, the mixing is described by a
$3\times 3$ unitary matrix $V$, which is usually
parameterized~\cite{Schechter:1980gr} by three mixing angles
($\theta_{12}$, $\theta_{23}$, and $\theta_{13}$) and three CP
violating phases out of which one is the Dirac phase ($\delta$) and
the other two are the Majorana phases ($\rho$ and $\sigma$). In the
standard parameterization advocated by the Particle Data
Group~\cite{Nakamura:2010zzi} and in
Refs.~\cite{Fritzsch:2001ty,*Xing:2003ez}, the lepton mixing matrix
reads
\begin{widetext}
\begin{eqnarray}\label{eq:SP}
V = \left(\begin{matrix}c_{12} c_{13} & s_{12} c_{13} & s_{13}
e^{-{\rm i}\delta} \cr -s_{12} c_{23} - c_{12} s_{23} s_{13}e^{{\rm
i}\delta} & c_{12} c_{23} - s_{12} s_{23} s_{13}e^{{\rm i}\delta} &
s_{23} c_{13} \cr s_{12} s_{23} - c_{12} c_{23} s_{13}e^{{\rm
i}\delta} & -c_{12} s_{23} - s_{12} c_{23} s_{13}e^{{\rm i}\delta} &
c_{23} c_{13}
\end{matrix}\right) \left(\begin{matrix} e^{{\rm i}\rho} & 0 & 0
\cr 0 & e^{{\rm i}\sigma} & 0 \cr 0 & 0 & 1 \end{matrix}\right)  ,
\end{eqnarray}
\end{widetext}
where $s_{ij} \equiv \sin \theta_{ij}$ and $c_{ij} \equiv \cos
\theta_{ij}$ (for $ij = 12, 23, 13$). If neutrinos are Dirac
particles, the phases $\rho$ and $\sigma$ will be irrelevant and can
be rotated away through a redefinition of the neutrino fields. The
latest global analysis of current neutrino oscillation data yields
\cite{Fogli:2011qn}
\begin{eqnarray}\label{eq:bound}
31.0^\circ < & \theta_{12} & < 37.1^\circ \; , \nonumber \\
35.7^\circ < & \theta_{23} & < 53.1^\circ \; ,  \\
4.1^\circ < & \theta_{13} & < 12.9^\circ \; ,\nonumber
\end{eqnarray}
at $3\sigma$ C.L., and the best-fit values of three mixing angles
are $\theta_{12} = 34.0^\circ$, $\theta_{23} = 40.4^\circ$ and
$\theta_{13} = 9.1^\circ$. Driven in particular by the latest T2K
results~\cite{Abe:2011sj}, $\theta_{13} = 0^\circ$ is currently
disfavored at the more than $3\sigma$ level.

So far it is still unclear how to theoretically understand the
observed lepton mixing. One tentative way is to start with
experimental values of leptonic mixing angles and conjecture a
simple constant mixing pattern, which may turn out to be suggestive
of the underlying symmetry of lepton mixing. In fact, several
interesting constant mixing patterns have been suggested along with
the progress in neutrino oscillation experiments, and shown to be
derivable from the flavor symmetries. For instance, the
democratic~\cite{Fritzsch:1995dj,*Fritzsch:1998xs},
bi-maximal~\cite{Barger:1998ta,*Vissani:1997pa},
tri-bimaximal~\cite{Harrison:2002er,*Xing:2002sw,*Harrison:2002kp,*He:2003rm},
hexagonal~\cite{Albright:2010ap,Kim:2010zub} or both golden
ratio~\cite{Kajiyama:2007gx,*Rodejohann:2008ir,*Everett:2008et,*Adulpravitchai:2009bg,*Feruglio:2011qq}
patterns can be realized in models with different discrete flavor
symmetries, such as $S_3$, $A_4$, $S_4$, $A_5$, dihedral groups,
etc., see \cite{Altarelli:2010gt,*Ishimori:2011nv} for recent
reviews, and \cite{Albright:2010ap} for a summary of proposed mixing
scenarios.

Note that all the aforementioned constant mixing patterns lead to a
vanishing $\theta_{13}$, which mainly due to long-baseline data is
now disfavored at $3\sigma$ C.L.~\cite{Fogli:2011qn}. While more
statistics and complementary measurements from reactor neutrino
experiments will tell us whether $\theta_{13}$ is indeed as large as
it currently appears to be, it is without doubt timely to consider
the ways to cope with a sizable $\theta_{13}$. Indeed, after the
results of T2K~\cite{Abe:2011sj} were released, several possible
ways, with a large range in what regards the level of
sophistication, to realize a relatively large $\theta_{13}$ have
been discussed in
Refs.~\cite{Xing:2011at,*He:2011gb,*Ma:2011yi,*Zhou:2011nu,
*Araki:2011wn,*Haba:2011nv,*Meloni:2011fx,*Morisi:2011wn,*Chao:2011sp,*Zhang:2011aw,*Chu:2011jg,*Dev:2011gi,*Toorop:2011jn,*Antusch:2011qg}.
Earlier analyses can be found for instance in
Refs.~\cite{Goswami:2009yy,King:2009qt,*Xing:2010pn,*He:2011kn,*Shimizu:2011xg}.

In this work, we search in a systematic way for successful mixing
patterns with initial nonzero $\theta_{13}$. Simple requirements are
the starting point of our analysis: (i) the mixing matrix is a
product of three rotations; (ii) the angles associated with the
rotations are simple fractions of $\pi$, such that the resulting
sines and cosines are given by exact expressions. In particular the
latter criterion is reminiscent of the results of many flavor
symmetry models. We find in total twenty viable mixing patterns in
the CP conserving case, while fifteen different feasible patterns
exist in the CP violating case. Furthermore, leaving the CP phase
arbitrary gives in total 66 successful mixing patterns.

The precision era which neutrino physics has recently entered
requires that the renormalization effects should be taken into
account. If the underlying flavor symmetry works at some high-energy
scale, the mixing angles will receive radiative corrections and
deviate from the predictions of a given mixing pattern when running
from the symmetry scale to the low-energy scale. We will therefore
consider the renormalization group equation (RGE) effects on our
scenarios.

The remaining part of this work is organized as follows. In
Sec.~\ref{sec:params}, we recall all different parameterizations of
a $3\times 3$ unitary matrix, and fix our notations. Then, in
Sec.~\ref{sec:patterns}, we summarize the feasible mixing patterns
which are compatible with experimental data and in particular
predict a nonzero $\theta_{13}$. Sec.~\ref{sec:RGE} is devoted to a
general discourse on the radiative corrections to the constant
mixing patterns. Finally, we conclude in Sec.~\ref{sec:summary}.

\section{Parameterizations of lepton mixing matrix}
\label{sec:params}

First of all, let us review the classification of possible
parameterizations of a lepton mixing matrix \cite{Fritzsch:1997st}.
Since the Majorana phases can always be recast into a diagonal
matrix on the right-hand side of $V$, they have no influence on our
results and will be ignored for now. If the leptonic CP violation is
absent, $V$ is simply a $3\times 3$ orthogonal matrix and can be
written as a product of three rotation matrices with three different
rotation angles $(\vartheta_1, \vartheta_2, \vartheta_3)$, i.e.
\begin{eqnarray}\label{eq:VR}
V=R_{ij}(\vartheta_1) R_{kl}(\vartheta_2) R_{mn}(\vartheta_3) \; ,
\end{eqnarray}
where $ij, kl, mn = 12, 23, 13$ and
\begin{eqnarray}
R_{12}(\vartheta) & = & \left(\begin{matrix}\cos\vartheta &
\sin\vartheta & 0 \cr -\sin\vartheta & \cos\vartheta & 0 \cr 0 & 0 &
1 \end{matrix}\right)  ,\nonumber \\
R_{23}(\vartheta) & = & \left(\begin{matrix}1 & 0 & 0 \cr 0 &
\cos\vartheta & \sin\vartheta \cr 0 & -\sin\vartheta & \cos\vartheta
\end{matrix}\right)  , \\
R_{13}(\vartheta) & = & \left(\begin{matrix}\cos\vartheta & 0 &
\sin\vartheta \cr 0 & 1 & 0 \cr -\sin\vartheta & 0 & \cos\vartheta
\end{matrix}\right) \nonumber .
\end{eqnarray}
Note that the order of the rotations is not specified. The
Dirac-type CP violating phase $\varphi$ can be included in the above
parameterization by replacing the entry ``1'' with a phase factor
$e^{-{\rm i}\varphi}$ in the second rotation matrix on the
right-hand side of Eq.~\eqref{eq:VR}. Taking $R_{12}(\vartheta)$ in
Eq. (4) for example, we have
\begin{eqnarray}\label{eq:phase}
R_{12}(\vartheta,\varphi) & = & \left(\begin{matrix}\cos\vartheta &
\sin\vartheta & 0 \cr -\sin\vartheta & \cos\vartheta & 0 \cr 0 & 0 &
e^{-{\rm i}\varphi}\end{matrix}\right)  .
\end{eqnarray}
Although there are several different ways to introduce the
CP violating phase, the choice in Eq.~\eqref{eq:phase} is
advantageous in the sense that the phase parameter $\varphi$ is
always located in a $2\times 2$ submatrix of $V$, in which each
element is a sum of two terms with the relative phase $\varphi$.

In Ref.~\cite{Fritzsch:1997st} it was shown that only nine distinct
parameterizations exist, namely
\begin{eqnarray}\label{eq:P}
P_1: ~~V &=& R_{12}(\vartheta_1) R_{23}(\vartheta_2,
\varphi) R^{-1}_{12}(\vartheta_3) \; , \nonumber \\
P_2: ~~ V &=& R_{23}(\vartheta_1) R_{12}(\vartheta_2,
\varphi) R^{-1}_{23}(\vartheta_3) \; , \nonumber \\
P_3: ~~ V &=& R_{23}(\vartheta_1) R_{13}(\vartheta_2,
\varphi) R_{12}(\vartheta_3) \; , \nonumber \\
P_4: ~~ V &=& R_{12}(\vartheta_1) R_{13}(\vartheta_2,
\varphi) R^{-1}_{23}(\vartheta_3) \; , \nonumber \\
P_5: ~~ V &=& R_{13}(\vartheta_1) R_{12}(\vartheta_2,
\varphi) R^{-1}_{13}(\vartheta_3) \; ,  \\
P_6: ~~ V &=& R_{12}(\vartheta_1) R_{23}(\vartheta_2,
\varphi) R_{13}(\vartheta_3) \; , \nonumber \\
P_7: ~~ V &=& R_{23}(\vartheta_1) R_{12}(\vartheta_2,
\varphi) R^{-1}_{13}(\vartheta_3) \; , \nonumber \\
P_8: ~~ V &=& R_{13}(\vartheta_1) R_{12}(\vartheta_2,
\varphi) R_{23}(\vartheta_3) \; , \nonumber \\
P_9: ~~ V &=& R_{13}(\vartheta_1) R_{23}(\vartheta_2, \varphi)
R^{-1}_{12}(\vartheta_3) \nonumber \; .
\end{eqnarray}
Here $R^{-1}_{ij}(\vartheta)=R_{ij}(-\vartheta)$. Three of the nine
parameterizations belong to the class $ij = mn \neq kl$ and six to
the class $ij \neq kl \neq mn$. Note that $P_3$ is just the standard
(``PDG'') parameterization in Eq.~\eqref{eq:SP}, up to a simple
phase redefinition.

The effects of CP violation are usually characterized by the Jarlskog invariant
$J_{\rm CP}$~\cite{Jarlskog:1985ht,*Wu:1985ea}, which is defined as
\begin{eqnarray}\label{eq:Jcp}
{\rm Im}\left[V_{\alpha i}V_{\beta j} V^*_{\alpha j}V^*_{\beta
i}\right] = J_{\rm CP} \sum_{\gamma = e,\mu,\tau} \sum_{k=1,2,3}
\left(\varepsilon_{ijk} \varepsilon_{\alpha \beta \gamma}\right) .
\end{eqnarray}
It is straightforward to verify that the Jarlskog invariant is given
by
\begin{eqnarray}
J_{\rm CP} = s_1 c_1 s^2_2 c_2 s_3 c_3 \sin\varphi \; ,
\end{eqnarray}
for $P_1, P_2$ and $P_5$, and
\begin{eqnarray}
J_{\rm CP} = s_1 c_1 s_2 c^2_2 s_3 c_3 \sin\varphi \; ,
\end{eqnarray}
for $P_3,P_4,P_6,P_7,P_8$ and $P_9$, where $s_i \equiv \sin
\vartheta_i$ and $c_i \equiv \cos \vartheta_i$ for $i = 1, 2, 3$.

It is worthwhile to remark that although these nine
parameterizations are mathematically equivalent, one of them may
turn out to be more useful than the others for a specific problem.
For instance, three mixing angles in the standard parameterization
$P_3$ can be unambiguously extracted from neutrino oscillation
experiments, which is not the case for the other parameterizations.
The flavor symmetry behind the observed lepton mixing pattern may be
manifest in a certain parameterization, which is currently unknown
to us, so we consider all possibilities in Eq.~\eqref{eq:P} when
searching for viable constant lepton mixing patterns.

\section{Simple rotation angles and viable mixing patterns}
\label{sec:patterns}

One immediate question arises: which values of the rotation angles
should we use to account for the observed lepton mixing? To make the
mixing patterns simple and suggestive of flavor symmetries, we set
the following criteria: (i) three rotation angles are simple
fractions of $\pi$, i.e.~$\pi/n$ with $n$ being an integer; (ii) the
choice of $n$ is governed by the requirement that the sines and
cosines of the rotation angles possess exact expressions, i.e.~they
are expressible as simple terms involving only square roots.
Although these two criteria are essentially set to make the lepton
mixing matrix as simple as possible, they might be realized in the
flavor symmetry models (e.g., the dihedral group
$D_n$~\cite{Grimus:2003kq,*Grimus:2004rj,*Blum:2007jz,*Blum:2007nt,*Ishimori:2008ns}),
where the lepton mixing patterns can be intimately related to
Clebsch--Gordan coefficients favoring small integers and their
square roots.
\begin{table}[t]
\renewcommand\arraystretch{1.8}\vspace{0.05cm}
\begin{center}
\begin{tabular}{|c|c|c|c|c|c|c|c|c|c|}
\hline $\displaystyle \vartheta = \frac{\pi}{n}$ & $\pi$ &
$\displaystyle \frac{\pi}{2}$ & $\displaystyle \frac{\pi}{3}$ &
$\displaystyle \frac{\pi}{4}$ & $\displaystyle \frac{\pi}{5}$ &
$\displaystyle \frac{\pi}{6}$ & $\displaystyle \frac{\pi}{8}$ &
$\displaystyle \frac{\pi}{10}$ & $\displaystyle \frac{\pi}{12}$\\
\hline $\sin^2 \vartheta$ & $~0$ & $~1$ & $\displaystyle
~\frac{3}{4}$ & $\displaystyle ~\frac{1}{2}$ & $\displaystyle
\frac{5-\sqrt{5}}{8}$ & $\displaystyle ~\frac{1}{4}$ &
$\displaystyle \frac{2-\sqrt{2}}{4}$ & $\displaystyle
\frac{3-\sqrt{5}}{8}$ & $\displaystyle \frac{2-\sqrt{3}}{4}$ \\
\hline $\cos^2 \vartheta$ & $~1$ & $~0$ & $\displaystyle
~\frac{1}{4}$ & $\displaystyle ~\frac{1}{2}$ & $\displaystyle
\frac{3+\sqrt{5}}{8}$ & $\displaystyle ~\frac{3}{4}~$ &
$\displaystyle \frac{2+\sqrt{2}}{4}$ & $\displaystyle
\frac{5+\sqrt{5}}{8}$ & $\displaystyle \frac{2+\sqrt{3}}{4}$ \\
\hline
\end{tabular}
\end{center}
\caption{\label{tab:sin} Exact expressions of $\sin^2 \vartheta$ and
$\cos^2 \vartheta$ for simple rotation angles $\vartheta = \pi/n$
with $n = 1, 2, 3, 4, 5, 6, 8, 10, 12$.}
\end{table}

Up to $n=12$, one has the following nine rotation angles satisfying
the above criteria
\begin{eqnarray}
\vartheta \in \left\{\pi, \frac{\pi}{2}, \frac{\pi}{3},
\frac{\pi}{4}, \frac{\pi}{5}, \frac{\pi}{6}, \frac{\pi}{8},
\frac{\pi}{10}, \frac{\pi}{12}\right\} ,
\end{eqnarray}
for which the values of $\sin^2 \vartheta$ and $\cos^2 \vartheta$
have been listed in Table~\ref{tab:sin}. Note that for $n>12$, there
also exist simple angles whose sines and cosines have exact
expressions. In general, one can always implement the relation
\begin{eqnarray}
\cos\frac{\vartheta}{2} =\sqrt{\frac{1-\cos\vartheta}{2}} \; ,
\end{eqnarray}
to calculate $\cos(\pi/2n)$ if $\cos (\pi/n)$ is already known.
However, it should be rather difficult to relate the expressions
with three or more square roots to a flavor symmetry, and in
addition the resultant lepton mixing pattern becomes very
complicated. Furthermore, as $n$ increases, $\vartheta$ becomes
smaller, and because the smallest mixing matrix element $V_{e3}$ is
generally related to the smallest rotation angle, the resulting
mixing matrix ends up not being in agreement with experimental data.

The lepton mixing matrix $V$ can be obtained by inserting the
rotation angles $\vartheta_i = \pi/n_i$ (for $i=1,2,3$) into
Eq.~\eqref{eq:P}, where $n_i \in \{1,2,3,4,5,6,8,10,12\}$. For later
convenience, we denote the mixing pattern constructed through the
parameterization $P_j$ (for $j = 1,2,\ldots,9$) with rotation angles
$\vartheta_i = \pi/n_i$ as $V = P_j(n_1,n_2,n_3)$ for a given CP
violating phase $\varphi$. Comparing between the standard
parameterization defined in Eq.~\eqref{eq:SP} and
$P_j(n_1,n_2,n_3)$, one can immediately extract the three
``standard'' leptonic mixing angles
\begin{eqnarray}
\sin^2\theta_{12} & = & \frac{|V_{e2}|^2}{1-|V_{e3}|^2} \; ,
\nonumber \\
\sin^2\theta_{23} & = & \frac{|V_{\mu 3}|^2}{1-|V_{e3}|^2} \; ,
 \\
\sin^2\theta_{13} & = & |V_{e3}|^2 \nonumber \; ,
\end{eqnarray}
which allow us to confront the lepton mixing matrix
$P_j(n_1,n_2,n_3)$ with neutrino oscillation data and thus find out
the viable patterns. Note that with our set of rotation angles we
can not generate a viable pattern with the standard parameterization
$P_3$ in Eq.~(\ref{eq:SP}), because $|V_{e3}| = \sin \theta_{13} $
in this case and the smallest possible angle in our scenario is
$\pi/12$, implying a too large value of $\sin^2 \theta_{13} = 0.07$.
The largest allowed value of $ \theta_{13} $ approximates to
$\pi/14$, which does not possess an exact expression for its sine or
cosine.

There exist nine distinct parameterization schemes and $9\times
9\times 9$ different combinations of rotation angles in each scheme,
so we are left with $9^4 = 6561$ possible mixing matrices for a
fixed CP violating phase $\varphi$. We have numerically studied all
these possibilities for the CP conserving case $\varphi=0$ and the
maximal CP violating case $\varphi=\pi/2$ by comparing the predicted
mixing angles with the $3\sigma$ global-fit data given in
Eq.~\eqref{eq:bound}. The viable cases are listed in
Table~\ref{tab:list}.
\begin{table}[t]
\vspace{0.1cm}
\begin{center}
\begin{tabular}{|c|c|c|c|c|c|c|c|}
\hline \multirow{2}*{Patterns} &  \multicolumn{3}{|c|}{$\varphi=0$}
&\multicolumn{4}{|c|}{$\varphi=\pi/2$} \\ \cline{2-8}& $\theta_{12}$ & $\theta_{23}$ & $\theta_{13}$ & $\theta_{12}$ & $\theta_{23}$ & $\theta_{13}$ & $J_{\rm CP} ~[\%]$ \\
\hline
~$P_1(10,4,4)$~ & $32.1^\circ$  & {\color{black}$43.6^\circ$}  & $12.6^\circ$ &&&&\\
\hline
~$P_1(10,4,5)$~ &  &  &  & $36.9^\circ$ & {\color{black}$43.6^\circ$} & $12.6^\circ$ & 4.9 \\
\hline
~$P_1(10,4,6)$~ &  &  &  & $31.6^\circ$ & {\color{black}$43.6^\circ$} & $12.6^\circ$ & 4.5 \\
\hline
~$P_1(12,4,4)$~ & {\color{black}$34.3^\circ$} & {\color{black}$44.0^\circ$} & $10.5^\circ$ &&&& \\
\hline
~$P_1(12,4,5)$~ &  &  &  & $36.6^\circ$ & {\color{black}$44.0^\circ$} &  $10.5^\circ$ & 4.2 \\
\hline
~$P_1(12,4,6)$~ &  &  &  & $31.1^\circ$ & {\color{black}$44.0^\circ$} &  $10.5^\circ$ & 3.8 \\
\hline
~$P_2(3,5,10)$~ & {\color{black}$34.6^\circ$} & $45.3^\circ$ & $10.5^\circ$ &&&& \\
\hline
~$P_2(3,5,12)$~ & $35.1^\circ$ & $47.8^\circ$ & {\color{black}$8.8^\circ$} &&&& \\
\hline
~$P_2(4,5,10)$~ &  &  &  & {\color{black}$34.6^\circ$} & {\color{black}$45.0^\circ$} &  $10.5^\circ$ & 4.1 \\
\hline
~$P_2(4,5,12)$~ &  &  &  & $35.1^\circ$ & {\color{black}$45.0^\circ$} &  {\color{black}$8.8^\circ$} & 3.5 \\
\hline
~$P_2(5,5,10)$~ &  &  &  & {\color{black}$34.6^\circ$} & $37.2^\circ$ &  $10.5^\circ$ & 3.9 \\
\hline
~$P_2(5,5,12)$~ &  &  &  & $35.1^\circ$ & $36.8^\circ$ &  {\color{black}$8.8^\circ$} & 3.3 \\
\hline
~$P_4(6,8,4)$~ & $36.3^\circ$ & $48.9^\circ$ &  $6.8^\circ$ &&&& \\
\hline
~$P_4(6,10,4)$~ & {\color{black}$33.4^\circ$} & $47.0^\circ$ & {\color{black}$9.5^\circ$} &&&& \\
\hline
~$P_4(6,10,5)$~ & {\color{black}$34.3^\circ$} & {\color{black}$39.5^\circ$} & $4.4^\circ$ &&&& \\
\hline
~$P_4(6,12,4)$~& $31.5^\circ$ & $45.9^\circ$ & $11.2^\circ$ &&&& \\
\hline
~$P_4(6,12,5)$~ & $32.7^\circ$ & $38.1^\circ$ & $6.5^\circ$ &&&& \\
\hline
~$P_4(8,6,5)$~ & $36.0^\circ$ & {\color{black}$44.9^\circ$} & {\color{black}$8.6^\circ$} &&&& \\
\hline
~$P_4(8,6,6)$~ & $35.1^\circ$ & {\color{black}$39.9^\circ$} & $12.0^\circ$ &&&&\\
\hline
~$P_4(10,6,4)$~ & {\color{black}$34.0^\circ$} & $52.0^\circ$ &  $6.8^\circ$ &&&& \\
\hline
~$P_4(10,6,5)$~ & $32.7^\circ$ & $44.3^\circ$ &  $11.7^\circ$ &&&& \\
\hline
~$P_4(12,6,4)$~ & $32.1^\circ$ & $51.7^\circ$ &  {\color{black}$9.1^\circ$} &&&& \\
\hline
~$P_5(5,4,3)$~ & $35.7^\circ$ & {\color{black}$38.7^\circ$} &  $11.6^\circ$ &&&& \\
\hline
~$P_7(4,5,12)$~ &  &  &  & $36.9^\circ$ & {\color{black}$45.0^\circ$} &  $12.1^\circ$ & 4.8 \\
\hline
~$P_7(5,5,12)$~ & $36.9^\circ$ & {\color{black}$45.0^\circ$} & $12.1^\circ$ & $36.9^\circ$ & $36.4^\circ$ &  $12.1^\circ$ & 4.6\\
\hline
~$P_7(6,5,12)$~ & $36.9^\circ$ & $39.0^\circ$ & $12.1^\circ$ &&&& \\
\hline
~$P_9(10,4,5)$~ &  &  &  & $36.9^\circ$ & $46.4^\circ$ &  $12.6^\circ$ & 4.9 \\
\hline
~$P_9(10,4,6)$~ &  &  &  & $31.6^\circ$ & $46.4^\circ$ &  $12.6^\circ$ & 4.5 \\
\hline
~$P_9(10,4,8)$~ & $35.4^\circ$ & $46.4^\circ$ & $12.6^\circ$ &&&& \\
\hline
~$P_9(12,4,5)$~ &  &  &  & $36.6^\circ$ & $46.0^\circ$ &  $10.5^\circ$ & 4.2 \\
\hline
~$P_9(12,4,6)$~ &  &  &  & $31.1^\circ$ & $46.0^\circ$ &  $10.5^\circ$ & 3.8 \\
\hline
~$P_9(12,4,8)$~ & {\color{black}$33.2^\circ$} & $46.0^\circ$ & $10.5^\circ$ &&&& \\
\hline
~$P_9(12,5,5)$~ &  &  &  & $36.4^\circ$ & $36.9^\circ$ &  $12.1^\circ$ & 4.6 \\
\hline
~$P_9(12,5,8)$~ & $31.5^\circ$ & $36.9^\circ$ & $12.1^\circ$ &&&& \\
\hline
\end{tabular}
\end{center}\vspace{-0.2cm}
\caption{\label{tab:list} List of viable mixing patterns, in which
the leptonic mixing angles are within the $3\sigma$ ranges of the
global-fit data. In the case of $\varphi=\pi/2$, the Jarlskog
invariant $J_{\rm CP}$ is given in the last column.}\vspace{-0.2cm}
\end{table}
We find that 20 mixing patterns are allowed in the CP conserving
case, while only 15 patterns are compatible with experimental data
in the maximal CP violating case. All in all, the feasible patterns
for different CP violating phases are different. Only $P_7(5,5,12)$
is allowed for both $\varphi=0$ and $\varphi=\pi/2$, showing the
importance of the CP violating phase in searching for successful
lepton mixing patterns. We should note here an important difference
between the standard parameterization $P_3$ in Eq.~(\ref{eq:SP}) and
the others. In the standard parameterization, a choice of three
angles can directly be confronted with experiments, namely knowing
$|V_{e2}|$, $|V_{e3}|$ and $|V_{\mu 3}|$ fixes the mixing angles and
leaves the phase undetermined. Consider now $P_2$, whose explicit
form is
\begin{eqnarray}\label{eq:P2}
V = \left(\begin{matrix}
c_2 & s_2 c_3 & -s_2 s_3 \\
-c_1 s_2 & c_1 c_2 c_3 + s_1 s_3 e^{-{\rm i}\varphi} & - c_1 c_2 s_3 +
s_1 c_3 e^{-{\rm i}\varphi} \\
s_1 s_2 & -s_1 c_2 c_3 + c_1 s_3 e^{-{\rm i}\varphi} & s_1 c_2 s_3 +
c_1 c_3 e^{-{\rm i}\varphi}
\end{matrix}\right) \nonumber .
\end{eqnarray}
Since $|V_{e2}|$, $|V_{e3}|$ and $|V_{\mu 3}|$ are known
experimentally, two angles ($\vartheta_2$ and $\vartheta_3$), as
well as some combination of $\vartheta_1$ and the CP phase
$\varphi$, can be determined. On the other hand, choosing three
angles, as we have done in this paper, fixes the CP phase $\varphi$
at the same time. This is true for all the alternative
parameterizations $P_{1,2,4,5,6,7,8,9}$ of the lepton mixing matrix.

We continue by recommending a few simple but interesting mixing
patterns and discuss their implications for the leptonic mixing
angles and leptonic CP violation:

{\it (1) Pattern $P_1(12,4,4)$ with $\varphi = 0$} -- The lepton
mixing matrix takes the form
\begin{eqnarray}
V=\left(\begin{matrix} \frac{\sqrt{3} - 1}{4\sqrt{2}} +
\frac{\sqrt{3}+1}{4} &  \frac{\sqrt{3} - 1}{4\sqrt{2}}
-\frac{\sqrt{3}+1}{4} &   \frac{\sqrt{3}-1}{4}\cr \frac{\sqrt{3} +
1}{4\sqrt{2}}  -\frac{\sqrt{3}-1}{4}  & \frac{\sqrt{3} +
1}{4\sqrt{2}} + \frac{\sqrt{3}-1}{4}  & \frac{\sqrt{3}+1}{4} \cr
-\frac{1}{2} &  -\frac{1}{2} &
\frac{1}{\sqrt{2}}\end{matrix}\right),
\end{eqnarray}
which leads to
\begin{eqnarray}
\sin^2\theta_{13} &=& \frac{1}{8}\left(2-\sqrt{3}\right) \;,
\nonumber \\
\sin^2\theta_{23} &=& \frac{2+\sqrt{3}}{6+\sqrt{3}} \; ,
 \\
\sin^2\theta_{12} &=& \frac{1}{2} - \frac{\sqrt{2}}{6+\sqrt{3}} \;,\nonumber
\end{eqnarray}
or explicitly $\theta_{13} \approx 10.5^\circ$, $\theta_{23} \approx
44.0^\circ$ and $\theta_{12} \approx 34.3^\circ$. Such a mixing
pattern is in excellent agreement with neutrino oscillation data.
The predictions for $\theta_{12}$ and $\theta_{23}$ fall into the
$1\sigma$ ranges, while that for $\theta_{13}$ is even slightly
larger than the best-fit value but well within the $2\sigma$ range.

{\it (2) Pattern $P_4(12,6,4)$ with $\varphi = 0$} -- The lepton
mixing matrix takes the form
\begin{eqnarray}
V = \left(\begin{matrix} \frac{3+\sqrt{3}}{4\sqrt{2}} &
\frac{3\sqrt{3}-1}{8} & \frac{3-\sqrt{3}}{8} \cr
-\frac{3-\sqrt{3}}{4\sqrt{2}} & \frac{3+\sqrt{3}}{8}  &
-\frac{3\sqrt{3}+1}{8} \cr -\frac{1}{2} & \frac{\sqrt{3}}{2\sqrt{2}}
& \frac{\sqrt{3}}{2\sqrt{2}}
\end{matrix}\right) ,
\end{eqnarray}
which leads to
\begin{eqnarray}
\sin^2\theta_{13} &=& \frac{3}{32}\left(2-\sqrt{3}\right) \;,
\nonumber \\
\sin^2\theta_{23} &=& \frac{14+3\sqrt{3}}{26+3\sqrt{3}} \; ,
\\
\sin^2\theta_{12} &=& \frac{14-3\sqrt{3}}{26+3\sqrt{3}} \;,\nonumber
\end{eqnarray}
or explicitly $\theta_{13} \approx 9.1^\circ$, $\theta_{23} \approx
51.7^\circ$ and $\theta_{12} \approx 32.1^\circ$. Note that the
prediction for $\theta_{13}$ in this mixing pattern is almost the
best-fit value.

{\it (3) Pattern $P_1(12,4,6)$ with $\varphi = \pi/2$} -- The
leptonic mixing matrix takes the form
\begin{eqnarray}
V = \left(\begin{matrix} \frac{\sqrt{3}-1- {\rm
i}\sqrt{2}(3+\sqrt{3})}{8} & \frac{3-\sqrt{3} + {\rm
i}\sqrt{2}(\sqrt{3}+1)}{8} & \frac{\sqrt{3}-1}{4} \cr
\frac{\sqrt{3}+1 + {\rm i}\sqrt{2}(3-\sqrt{3})}{8} &
\frac{3+\sqrt{3} - {\rm i}\sqrt{2}(\sqrt{3}-1)}{8}   &
\frac{\sqrt{3}+1}{4} \cr -\frac{1}{2\sqrt{2}} &
-\frac{\sqrt{3}}{2\sqrt{2}} & \frac{1}{\sqrt{2}}
\end{matrix}\right) ,
\end{eqnarray}
which leads to
\begin{eqnarray}
\sin^2\theta_{13} &=& \frac{1}{8}\left(2-\sqrt{3}\right) \;,
\nonumber \\
\sin^2\theta_{23} &=& \frac{2+\sqrt{3}}{6+\sqrt{3}} \; ,
 \\
\sin^2\theta_{12} &=& \frac{10-\sqrt{3}}{24+4\sqrt{3}}\;,\nonumber
\end{eqnarray}
or explicitly $\theta_{13} \approx 10.5^\circ$, $\theta_{23} \approx
44.0^\circ$ and $\theta_{12} \approx 31.1^\circ$. In addition, the
Jarlskog invariant is $J_{\rm CP} = \sqrt{6}/64 \approx 3.8\%$.

{\it (4) Pattern $P_9(12,4,6)$ with $\varphi = \pi/2$} -- The lepton
mixing matrix takes the form
\begin{eqnarray}
V = \left(\begin{matrix} \frac{-\sqrt{3}+1 - {\rm
i}\sqrt{2}(3+\sqrt{3})}{8} & \frac{-3+\sqrt{3} + {\rm
i}\sqrt{2}(\sqrt{3}+1)}{8} & \frac{\sqrt{3}-1}{4} \cr
\frac{1}{2\sqrt{2}} & \frac{\sqrt{3}}{2\sqrt{2}} &
\frac{1}{\sqrt{2}} \cr \frac{-\sqrt{3}-1 + {\rm
i}\sqrt{2}(3-\sqrt{3})}{8} & \frac{-3-\sqrt{3} - {\rm
i}\sqrt{2}(\sqrt{3}-1)}{8}   & \frac{\sqrt{3}+1}{4}
\end{matrix}\right) , \nonumber \\
\end{eqnarray}
which leads to
\begin{eqnarray}
\sin^2\theta_{13} &=& \frac{1}{8}\left(2-\sqrt{3}\right) \;,
\nonumber \\
\sin^2\theta_{23} &=& \frac{4}{6+\sqrt{3}} \; ,
\\
\sin^2\theta_{12} &=& \frac{10-\sqrt{3}}{24+4\sqrt{3}}\;,\nonumber
\end{eqnarray}
or explicitly $\theta_{13} \approx 10.5^\circ$, $\theta_{23} \approx
46.0^\circ$ and $\theta_{12} \approx 31.1^\circ$. In addition, the
Jarlskog invariant is $J_{\rm CP} = \sqrt{6}/64 \approx 3.8\%$,
which is the same as that in the previous case. Note that the mixing
patterns in Eqs.~(17) and (19) differ only in the predictions of
$\theta_{23}$, i.e.~$\theta_{23} = 44.0^\circ$ in the former case
while $\theta_{23} = 46.0^\circ$ in the latter.

\begin{sidewaysfigure*}[ph]
\begin{center}\vspace{12cm}
\includegraphics[width=23cm,bb=20 -140 720 560]{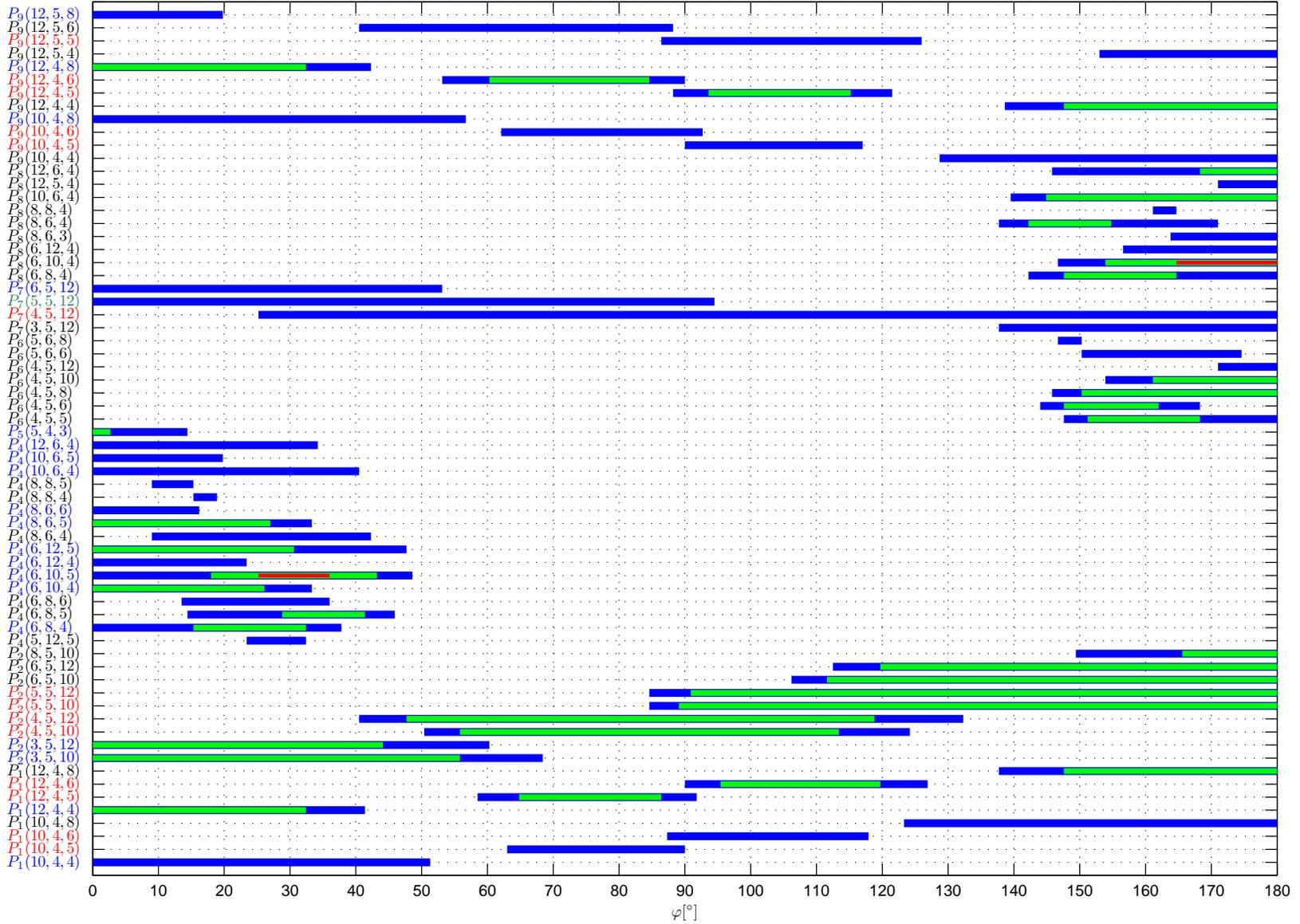}
\end{center}
\vspace{-11cm}
\parbox{6in}{\caption{\label{fig:pattern} Viable
patterns for an arbitrary CP violating phase $\varphi$. The red,
green, and blue bars denote the allowed ranges of $\varphi$, for
which the predicted mixing angles fall into their $1\sigma$,
$2\sigma$, and $3\sigma$ intervals, respectively. Patterns with
$\varphi=0$ are written in blue, while patterns with $\varphi=\pi/2$
are written in red. The pattern accommodating both $\varphi=0$ and
$\varphi=\pi/2$ is highlighted in green.}}
\end{sidewaysfigure*}

As mentioned before, the CP violating phase plays an important role
in searching for successful lepton mixing patterns. In this regard,
we have so far focused on the special cases with $\varphi = 0$ and
$\varphi = \pi/2$. Choosing an arbitrary CP violating phase,
i.e.~$\varphi \in [0, \pi]$, leads to the 66 viable patterns in
Fig.~\ref{fig:pattern}. One can immediately construct the lepton
mixing matrix from the $P_j(n_1,n_2,n_3)$ notation. It is amazing
that from the 6561 possible mixing patterns, only about $1\%$ are
compatible with current oscillation data.

\section{Radiative corrections}
\label{sec:RGE}

Now we proceed to consider possible radiative corrections to the
mixing patterns in Table~\ref{tab:list}.\footnote{In the flavor
symmetry models, the corrections to fermion masses and mixing
patterns may also originate from the flavor symmetry breaking and
the higher-dimensional operators. Since the significance of these
potential corrections is highly model-dependent, we shall
concentrate on the generic RGE corrections in the current study.} As
argued in Sec.~\ref{sec:intro}, the mixing patterns under
consideration may arise from certain flavor symmetries preserved at
high-energy scales, such as the grand unification (e.g.~$\Lambda
\sim 10^{16}~{\rm GeV}$) or the seesaw scale (e.g.~$\Lambda \sim
10^{14}~{\rm GeV}$), whereas the leptonic mixing parameters are
determined or constrained in neutrino oscillation experiments at low
energies. The gap between the high-energy predictions and the
low-energy measurements is bridged by the RG evolution, which may
significantly change the model predictions. On the other hand, the
RG running effects could also serve as an explanation for the
discrepancy between the flavor symmetric mixing pattern and the
observed one.

The RGEs for leptonic mixing parameters have been derived within
various theoretical
frameworks~\cite{Chankowski:1993tx,*Babu:1993qv,*Antusch:2001ck,
*Antusch:2001vn,*Chao:2006ye,*Schmidt:2007nq,*Chakrabortty:2008zh,
*Blennow:2011mp}.
In the supersymmetric theories with large $\tan\beta$, it has been
found that the RG evolution may lead to significant modifications to
the mixing parameters, in particular the solar mixing angle
$\theta_{12}$~(see e.g.~Ref.~\cite{Ray:2010rz} and references
therein). To be explicit, we write down the RGEs for three leptonic
mixing angles in the approximation of $\tau$-lepton dominance (i.e.,
$Y^{}_\ell Y^\dagger_\ell \approx {\rm diag} \{0,0,y^2_\tau\}$ in
view of $y^2_e \ll y^2_\mu \ll y^2_\tau $~~\cite{Antusch:2003kp}),
\begin{eqnarray}\label{eq:RGE}
\Dot{\theta}_{12} & \approx & -\frac{Cy_\tau^2 s^2_{12} c^2_{12}
s_{23}^2 }{8\pi^2 \Delta m^2_{\rm sol}} \left[m^2_1 + m^2_2 + 2 m_1
m_2 c_{2(\rho - \sigma)}\right] \; ,
\nonumber \\
\Dot{\theta}_{13} & \approx & + \frac{Cy_\tau^2 s^2_{12}c^2_{12}
s^2_{23} c^2_{23} m_3}{2\pi^2\Delta m^2_{\rm
atm}\left(1+\zeta\right)} \times \left[ m_1 c_{(2\rho+\delta)}
\right. \nonumber
\\ & & \left. ~~~~~~~ - \left(1+\zeta\right) m_2 c_{(2\sigma+\delta)} - \zeta m_3
c_\delta \right] \; ,
 \\
\Dot{\theta}_{23} & \approx & -\frac{Cy_\tau^2 s^2_{23} c^2_{23}
}{8\pi^2\Delta m^2_{\rm atm}} \times \left[c_{12}^2 \left(m^2_2 +
m^2_3 +
2m_2 m_3 c_{2\sigma}\right) \right. ~~~~~\nonumber \\
&& \left. + s_{12}^2 \left(m^2_1 + m^2_3 + 2 m_1 m_3
c_{2\rho}\right)(1+\zeta)^{-1} \right] \nonumber \; ,
\end{eqnarray}
where $\Dot{\theta}_{ij} \equiv {\rm d}\theta_{ij}/{\rm d}t$ with $t
= \ln (\mu/\mu_0)$, $\zeta \equiv \Delta m^2_{\rm sol}/\Delta
m^2_{\rm atm}$ with $\Delta m^2_{\rm sol} \equiv m^2_2 - m^2_1
\approx 7.6\times 10^{-5}~{\rm eV}^2$ and $|\Delta m^2_{\rm atm}|
\equiv |m^2_3 - m^2_2| \approx 2.3 \times 10^{-3}~{\rm eV}^2$ at the
low-energy scale, and $y_\tau$ denotes the Yukawa coupling of tau
lepton. In the standard model (SM) $C=-3/2$ while in the minimal
supersymmetric standard model (MSSM) $C=1$. Terms of ${\cal
O}(\theta_{13})$ have been safely neglected in Eq.~(\ref{eq:RGE}),
where we have defined $c_{2(\rho - \sigma)} \equiv \cos 2(\rho -
\sigma)$, $c_{2\rho} \equiv \cos 2\rho$ and so on.

As obvious from the above beta functions of the mixing angles,
$\theta_{12}$ receives typically larger RG corrections than
$\theta_{23}$ and $\theta_{13}$, whose corrections are of the same
order. Furthermore, when running from a high-energy scale to the
electroweak scale $\Lambda_{\rm EW} = 10^{2}~{\rm GeV}$, the
radiative corrections to $\theta_{12}$ are typically (see below)
positive in the MSSM, i.e.~$\theta_{12}(\Lambda) <
\theta_{12}(\Lambda_{\rm EW})$. In contrast, in the SM $\theta_{12}$
receives only negative corrections because of the sign flip in $C$.
However, the RG effects in the SM are generally small due to the
absence of $\tan \beta$ enhancement. As for $\theta_{23}$, the RG
corrections could be either positive or negative, depending mainly
on the model and the neutrino mass ordering. In the normal mass
ordering ($m_1<m_2<m_3$) both typically decrease in the SM and
increase in the MSSM, whereas for the inverted mass ordering
($m_3<m_1<m_2$) the behavior is opposite. Finally, nonzero
$\theta_{13}$ can run in both directions, and receives corrections
of the same order as $\theta_{23}$.

\begin{figure}[t]\vspace{-0.1cm}
\includegraphics[width=0.48\textwidth]{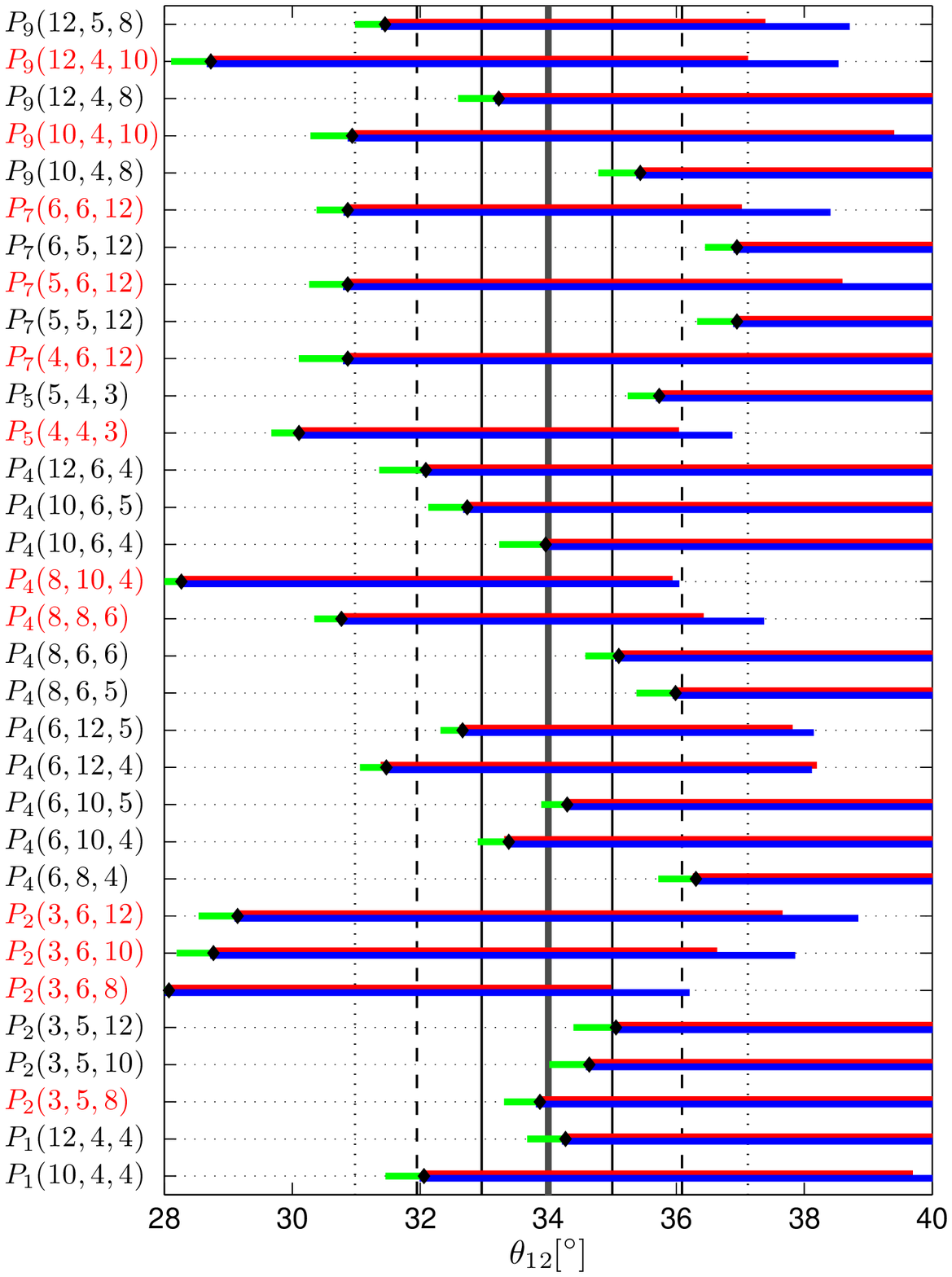}\vspace{-0.0cm}
\caption{\label{fig:phi0theta12}RG correction to $\theta_{12}$ in
the SM (green bars), and in the MSSM (red bars for the normal mass
ordering and blue bars for the inverted mass ordering). The initial
values are labeled as black diamonds. Note that $\varphi=0$ is
assumed and $\tan\beta=10$ is adopted in the MSSM. Furthermore, we
allow the lightest neutrino mass $m_1$ (or $m_3$) to vary in the
range of $(0 \ldots 0.1)~{\rm eV}$ in the MSSM, whereas $m_1$ (or
$m_3$) varies in the range of $(0\ldots 0.2)~{\rm eV}$ in the SM.
The vertical lines correspond to the best-fit value and the
$1\sigma$, $2\sigma$ and $3\sigma$ intervals. The mixing scenarios
written in red are not valid without RG corrections.}\vspace{-0.2cm}
\end{figure}
\begin{figure}[t]\vspace{-0.1cm}
\includegraphics[width=0.48\textwidth]{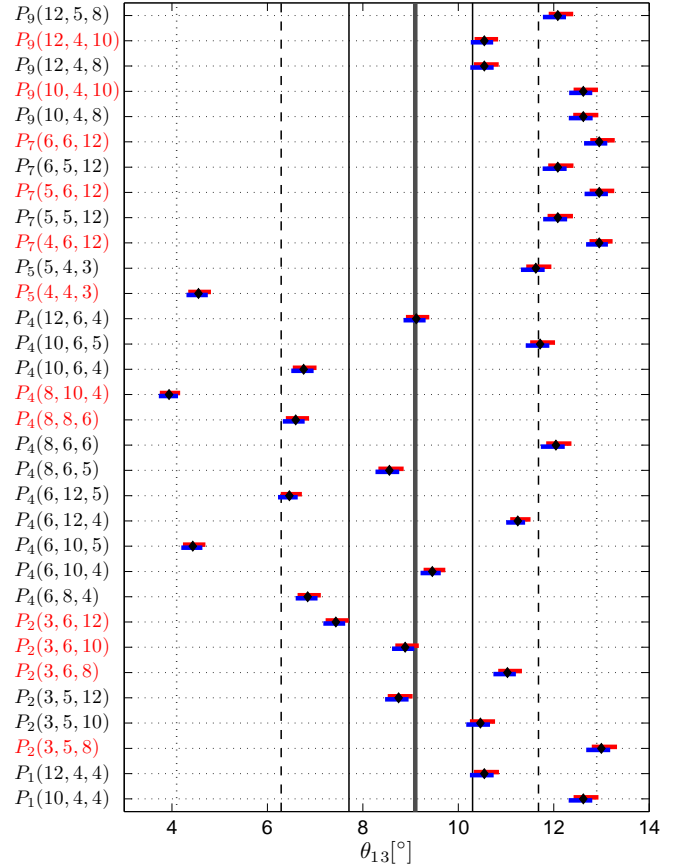}\vspace{-0.0cm}
\caption{\label{fig:phi0theta13} The RG corrections to $\theta_{13}$
in the SM (green bars), and in the MSSM (red bars for the normal
mass ordering and blue bars for the inverted mass ordering). The
initial values are labeled as black diamonds. The input parameters
are the same as in Fig.~\ref{fig:phi0theta12}. Note that the
corrections in the SM appear to be invisible. The mixing scenarios
written in red are not valid without RG corrections.}\vspace{-0.2cm}
\end{figure}
\begin{figure}[t]\vspace{-0.1cm}
\includegraphics[width=0.48\textwidth]{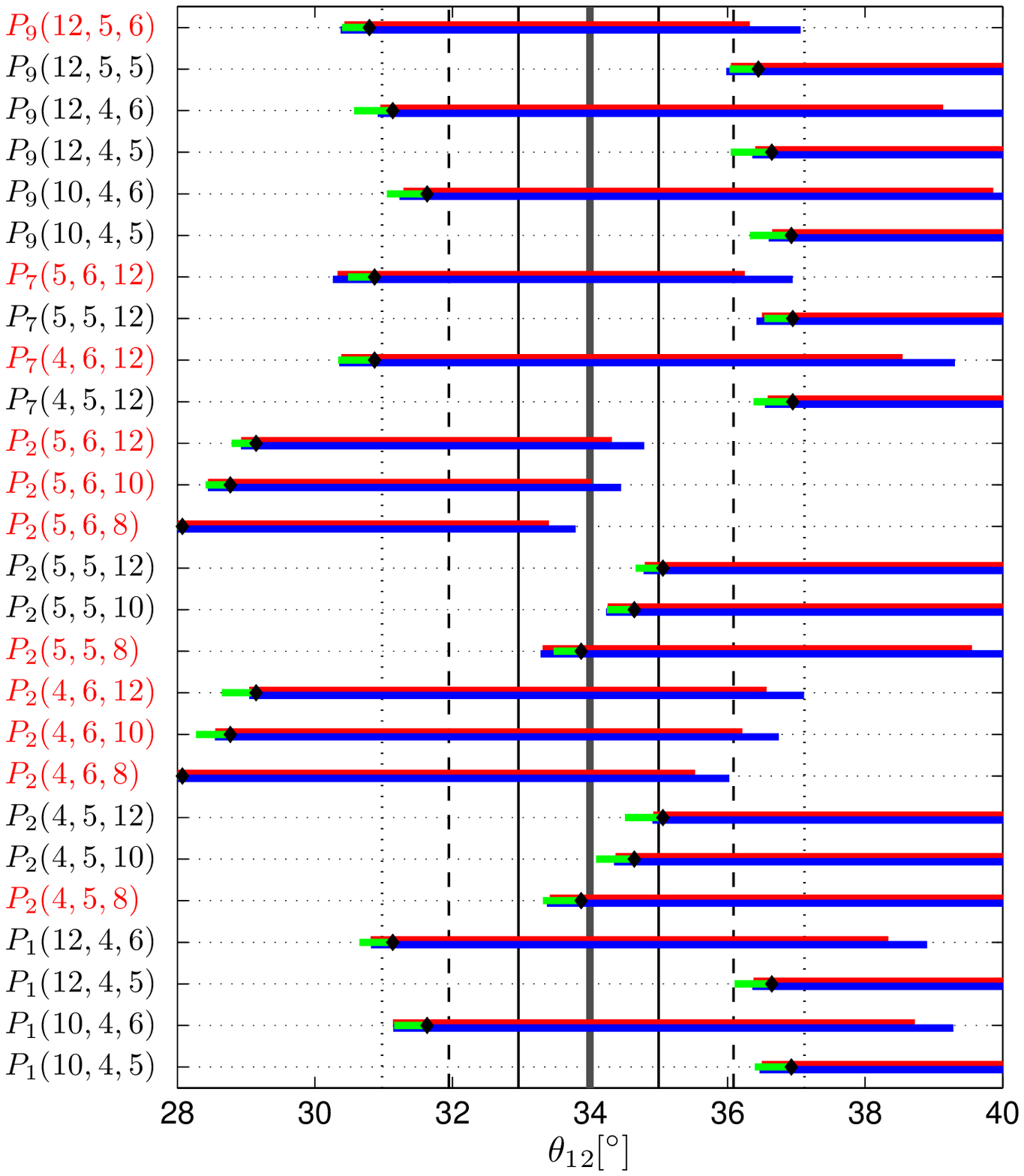}\vspace{-0.cm}
\caption{\label{fig:phi90theta12} The RG corrections to
$\theta_{12}$ in the SM (green bars), and in the MSSM (red bars for
the normal mass ordering and blue bars for the inverted mass
ordering). The initial values are labeled as black diamonds. We use
the same parameters as in Fig.~\ref{fig:phi0theta12} except
$\varphi=\pi/2$.} \vspace{-0.2cm}
\end{figure}
When searching for viable lepton mixing patterns, we have ignored
the Majorana phases, which indeed do not change the mixing angles.
However, they may play a significant role in the RG evolution of
mixing angles, in particular for $\theta_{12}$. In the CP conserving
limit with $\rho=\sigma=0$, $\Dot \theta_{12}\propto
(m_1+m_2)^2/\Delta m^2_{\rm sol}$, which may strongly boost the RG
running in the nearly degenerate or inverted mass ordering. On the
contrary, in the limit of $\rho-\sigma =\pi/2$, $\Dot
\theta_{12}\propto (m_2-m_1)/(m_2+m_1)$, which is always smaller
than one and hence suppresses the RG effects on $\theta_{12}$. In
the latter case, the running mainly comes from the
next-to-leading-order terms of $\theta_{13}$, and thus $\theta_{12}$
may run to a slightly smaller value. Since both $\rho$ and $\sigma$
are entirely unconstrained in oscillation experiments, we shall
allow them to freely vary between $0$ and $\pi$.

We have numerically solved the full set of RGEs for leptonic mixing
angles. More explicitly, the lepton mixing patterns in
Table~\ref{tab:list} are assumed at a cutoff scale $\Lambda= 10^{10}
~{\rm GeV}$, and then we evolve the mixing parameters down to the
electroweak scale $\Lambda_{\rm EW}$ in order to compare them with
experimental data. Note that the dependence of RG corrections on the
cutoff scale is logarithmic and therefore the precise value of
$\Lambda$ is not quite relevant. Other physical parameters,
e.g.~gauge couplings and fermion masses, are taken from
Ref.~\cite{Xing:2007fb}. In the case with $\varphi=0$, we show in
Fig.~\ref{fig:phi0theta12} the RG evolution of $\theta_{12}$ in the
MSSM with $\tan \beta = 10$ as well as in the SM. One can observe
that in the MSSM some viable mixing patterns can in principle
receive RG corrections so large that they are no longer valid, for
instance $P_9(10,4,8)$ in the inverted mass ordering. However, as
mentioned above, by choosing $\rho-\sigma =\pi/2$ one can suppress
the running of $\theta_{12}$, a situation with interesting
consequences for neutrinoless double beta decay, as discussed in
Ref.~\cite{Goswami:2009yy}. Note that, although the predictions on
$\theta_{12}$ are same for both $P_2(3,5,12)$ and $P_4(8,6,6)$, they
suffer from different RG corrections, reflecting the importance of
$\theta_{23}$ in the RG evolution [cf.~Eq.~\eqref{eq:RGE}]. In the
SM, the RG corrections are in general small, and all the viable
mixing patterns remain valid at the electroweak scale. In
Fig.~\ref{fig:phi0theta12} we have also shown twelve new mixing
scenarios, which generate too small or large $\theta_{12}$ when not
corrected by RG effects, but can enter the allowed range after RG
corrections. These cases are indicated by writing them in red (see
below).


For completeness, we also show the RG corrections to $\theta_{13}$
in Fig.~\ref{fig:phi0theta13}. As expected, no visible effects can
be seen in the SM, while in the MSSM, the RG running may lead to
tiny deviations (less than one degree) from their initial values.
Even in this case, the values of $\theta_{13}$ from some mixing
patterns listed in Table~\ref{tab:list} may exceed their $3\sigma$
ranges, e.g., the mixing pattern $P_1(10,4,4)$. Furthermore, the RG
effects on $\theta_{23}$ are of similar size to those on
$\theta_{23}$ [c.f. Eq.~\eqref{eq:RGE}], and therefore no
significant radiative corrections can be acquired.

Now we consider the maximal CP violating case with $\varphi=\pi/2$.
The RG corrections to $\theta_{12}$ are depicted in
Fig.~\ref{fig:phi90theta12}. Similar to the CP conserving case,
sizable RG effects can be present. In the MSSM and in the case of
the inverted mass ordering, one can observe that for instance
$P_9(12,5,5)$ can become incompatible with experimental data, and
similarly for $P_9(12,4,6)$ in the SM. On the other hand, there are
eleven new mixing schemes which become valid only after sizable RG
corrections.

Let us turn to the mixing patterns which are not compatible with
data at the cutoff scale but can be modified to the proper parameter
ranges with the help of the RG running. This is in particular
possible for the mixing patterns with both $\theta_{23}$ and
$\theta_{13}$ in the currently-favored ranges but a smaller
$\theta_{12}$, since $\theta_{12}$ may be lifted up into the correct
parameter interval in the MSSM. For example, the mixing patterns
with $\theta_{12}=30^\circ$, or $\sin\theta_{12} =1/2$, could be of
particular
interest~\cite{Albright:2010ap,Kim:2010zub}\footnote{Another example
of this kind is the so-called tetra-maximal mixing pattern with
$\theta_{12} \approx 30.4^\circ$ ($\tan \theta_{12} =
2-\sqrt{2}$)~\cite{Xing:2008ie}.}. We have also performed a
systematic search for such mixing patterns in our scenario. As
mentioned above, in the case of $\varphi = 0$ ($\varphi = \pi/2$),
there exist twelve (eleven) new patterns. Now we shortly discuss
three of them:

{\it (1) Pattern $P_5(4,4,3)$ with $\varphi = 0$} -- The lepton
mixing matrix is
\begin{eqnarray}
V=\left(\begin{matrix} \frac{\sqrt{6} + 1}{4} &  \frac{1}{2} &
\frac{\sqrt{2}-\sqrt{3}}{4}\cr \frac{\sqrt{2}}{4} &
\frac{\sqrt{2}}{2} & \frac{\sqrt{6}}{4} \cr \frac{\sqrt{6}-1}{4} &
-\frac{1}{2} & \frac{\sqrt{2}+\sqrt{3}}{4}\end{matrix}\right),
\end{eqnarray}
which leads to
\begin{eqnarray}
\sin^2\theta_{13} &=& \frac{1}{16}(5-2\sqrt{6}) \;,
\nonumber \\
\sin^2\theta_{23} &=& \frac{6}{11+2\sqrt{6}} \; ,
 \\
\sin^2\theta_{12} &=& \frac{4}{11+2\sqrt{6}} \;,\nonumber
\end{eqnarray}
or explicitly $\theta_{13} \approx 4.6^\circ$, $\theta_{23} \approx
38.0^\circ$ and $\theta_{12} \approx 30.1^\circ$. Evolving the
mixing angles through the full set of RGEs in the MSSM with
$\tan\beta=10$, we obtain, at the electroweak scale, $\theta_{12}
\in [30.1^\circ,~36.0^\circ]$ for the normal mass ordering and
$\theta_{12} \in [30.1^\circ,~36.9^\circ]$ for the inverted mass
ordering [cf.~Fig.~\ref{fig:phi0theta12}]. Thus the mixing angles
become consistent with the experimental data with the help of
radiative corrections.

{\it (2) Pattern $P_2(3,6,12)$ with $\varphi = 0$} -- The lepton
mixing matrix takes the form
\begin{eqnarray}
V = \left(\begin{matrix} \frac{\sqrt{3}}{2} &
\frac{\sqrt{3}+1}{4\sqrt{2}} & -\frac{\sqrt{3}-1}{4\sqrt{2}} \cr
-\frac{1}{4} & \frac{9-\sqrt{3}}{8\sqrt{2}}  &
\frac{3(\sqrt{2}+\sqrt{6})}{16} \cr \frac{\sqrt{3}}{4} &
-\frac{5+\sqrt{3}}{8\sqrt{2}} & \frac{5\sqrt{3}-1}{8\sqrt{2}}
\end{matrix}\right),
\end{eqnarray}
which leads to
\begin{eqnarray}
\sin^2\theta_{13} &=& \frac{1}{16}(2-\sqrt{3}) \;, \nonumber
\\
\sin^2\theta_{23} &=& \frac{9(2+\sqrt{3})}{4(14+\sqrt{3})}\; ,
 \\
\sin^2\theta_{12} &=& \frac{2+\sqrt{3}}{14+\sqrt{3}} \;, \nonumber
\end{eqnarray}
or explicitly $\theta_{13} \approx 7.4^\circ$, $\theta_{23} \approx
46.9^\circ$ and $\theta_{12} \approx 29.1^\circ$. Again we evolve
the mixing angles via the RGEs in the MSSM with $\tan\beta=10$, and
obtain $\theta_{12} \in [29.2^\circ,~37.7^\circ]$ in the normal mass
ordering case while $\theta_{12} \in [29.1^\circ,~38.8^\circ]$ in
the inverted mass ordering [cf.~Fig.~\ref{fig:phi0theta12}]. These
values of $\theta^{}_{12}$ are well compatible with the $3\sigma$
range in Eq. (2).

{\it (3) Pattern $P_2(4,6,12)$ with $\varphi = \pi/2$} -- The lepton
mixing matrix takes the form
\begin{eqnarray}
V = \left(\begin{matrix} \frac{\sqrt{3}}{2} & \frac{\sqrt{3} +
1}{4\sqrt{2}} & -\frac{\sqrt{3}-1}{4\sqrt{2}} \cr
-\frac{\sqrt{2}}{4} & \frac{3+\sqrt{3} - 2{\rm i}(\sqrt{3}-1)}{8}
& -\frac{3-\sqrt{3} + 2{\rm i}(\sqrt{3}+1)}{8} \cr
\frac{\sqrt{2}}{4}  & -\frac{3+\sqrt{3} + 2{\rm i}(\sqrt{3}-1)}{8}
& \frac{3-\sqrt{3} - 2{\rm i}(\sqrt{3}+1)}{8}
\end{matrix}\right) ,
\end{eqnarray}
which leads to
\begin{eqnarray}
\sin^2\theta_{13} &=& \frac{1}{16}(2-\sqrt{3}) \;,
\nonumber \\
\sin^2\theta_{23} &=& \frac{1}{2} \; ,
 \\
\sin^2\theta_{12} &=& \frac{2+\sqrt{3}}{14+\sqrt{3}} \;,\nonumber
\end{eqnarray}
or explicitly $\theta_{13} \approx 10.5^\circ$, $\theta_{23} =
45.0^\circ$ and $\theta_{12} \approx 29.1^\circ$. In addition, the
Jarlskog invariant is $J_{\rm CP} = \sqrt{3}/64 \approx 2.7\%$.
Similar to the above two patterns, the RG running in the MSSM leads
to $\theta_{12} \in [29.0^\circ,~36.6^\circ]$ for the normal mass
ordering and $\theta_{12} \in [29.0^\circ,~37.1^\circ]$ for the
inverted mass ordering [cf.~Fig.~\ref{fig:phi90theta12}], which are
in agreement with the experimental data.

In general, the evaluation of RGEs may play a crucial role in
searching for realistic mixing patterns. Especially in the
supersymmetric case, a larger $\tan\beta$ typically leads to more
significant RG corrections. In this sense, the mixing patterns with
smaller $\theta_{12}$ could be more favorable for a larger
$\tan\beta$. It is also worthwhile to stress that the RGEs under
discussion are given in the effective theory approach, which is
essentially the same for different kinds of flavor or seesaw models.
In the realistic flavor symmetry models, the flavons might induce
additional contributions to the RGEs. If a seesaw model is
considered, the RG running between the seesaw thresholds may also
lead to remarkable modifications, and thus should be carefully
treated~\cite{Antusch:2002rr,*Antusch:2005gp,*Bergstrom:2010id}. A
thorough survey on the RGEs in a specific model and on the different
values of $\tan\beta$ is beyond the scope of this work, and we refer
the readers to
Refs.~\cite{Chankowski:1993tx,Antusch:2003kp,Ray:2010rz} for more
detailed discussions.

\section{Conclusions}
\label{sec:summary}

Motivated by the recent indications of a nonzero $\theta_{13}$, we
have performed a systematic search for simple but viable lepton
mixing patterns by setting two criteria: (i) the lepton mixing
matrix is parameterized by three rotation angles, which are simple
fractions of $\pi$; (ii) the sines and cosines of these rotation
angles possess exact expressions. In total, we have found 20 viable
mixing patterns in the CP conserving limit, while 15 viable patterns
exist in case of maximal CP violation. Moreover, in the most general
cases with the CP phase unconstrained, only 66 mixing patterns out
of 6561 combinations are found to be compatible with current data.

Furthermore, radiative corrections to the mixing patterns have been
calculated by solving the RGEs of leptonic mixing parameters. We
have shown that the RG running can induce sizable corrections to the
lepton mixing patterns, which eventually could render some patterns
to be unsuccessful in describing lepton mixing. On the other hand,
we have also pointed out some interesting mixing patterns, which are
incompatible with current oscillation data at the high-energy scale
but become viable at the low-energy scale after the RG corrections
are properly taken into account.

We hope that the successful constant mixing patterns found in this
work can be helpful in searching for the underlying flavor
symmetries and shed some light on the final solution to the flavor
puzzle. At least, they could serve as a useful phenomenological
description of lepton mixing.

\begin{acknowledgments}
The authors would like to thank Zhi-zhong Xing for valuable
suggestions and partial involvement at the early stage of this work,
which was supported in part by the ERC under the Starting Grant
MANITOP and by the DFG in the Transregio 27 ``Neutrinos and Beyond''
(W.R. and H.Z.), and by the Alexander von Humboldt Foundation (S.Z.).
\end{acknowledgments}

\bibliography{bib}
\bibliographystyle{apsrevM}

\end{document}